\documentstyle[12pt,amssymb,amscd]{article}   

\def\AFOUR{%
\setlength{\textheight}{9.0in}%
\setlength{\textwidth}{5.75in}%
\setlength{\topmargin}{-0.375in}%
\hoffset=-.5in%
\renewcommand{\baselinestretch}{1.17}%
\setlength{\parskip}{6pt plus 2pt}%
}
\AFOUR                                           
\def\car{\mathop{\square}}
\def\carre#1#2{\raise 2pt\hbox{$\scriptstyle #1$}\car_{#2}}

\makeatletter
\def\section{\@startsection {section}{1}{\z@}{-3.5ex plus -1ex minus
 -.2ex}{2.3ex plus .2ex}{\large\bf}}
\def\subsection{\@startsection{subsection}{2}{\z@}{-3.25ex plus -1ex minus
 -.2ex}{1.5ex plus .2ex}{\normalsize\bf}}
\makeatother
\makeatletter
\@addtoreset{equation}{section}

\makeatother
\newcommand{\nc}{\newcommand}
\newcommand{\rnc}{\renewcommand}
\nc{\be}{\begin{equation}}
\nc{\ee}{\end{equation}}
\nc{\bea}{\begin{eqnarray}}
\nc{\eea}{\end{eqnarray}}

\def\slash#1{\setbox0=\hbox{$#1$}#1\hskip-\wd0\hbox to\wd0{\hss\sl/\/\hss}}

\def\href#1#2{{#2}}

\rnc{\a}{\alpha}
\nc{\ab}{\bar{\a}}
\nc{\ap}{\a^{+}}
\nc{\abm}{\ab^{-}}
\rnc{\b}{\beta}
\nc{\bb}{\bar{\b}}
\nc{\bbp}{\bb_{\zb}^{+}}
\nc{\bm}{\b_{z}^{-}}
\nc{\oa}{\overline{\a}}
\nc{\ob}{\overline{\b}}
\rnc{\gg}{\gamma}
\rnc{\d}{\delta}
\nc{\f}{\phi}
\nc{\fb}{\bar{\phi}}
\nc{\vf}{\varphi}
\nc{\p}{\psi}

\rnc{\c}{\chi}
\nc{\la}{\lambda}
\nc{\m}{{\mathrm m}}
\nc{\n}{\nu}
\rnc{\o}{\omega}
\nc{\Om}{\Omega}
\rnc{\t}{\theta}
\nc{\eps}{\epsilon}
\rnc{\S}{\Sigma}
\nc{\F}{\Phi}
\nc{\trac}[2]{{\textstyle\frac{#1}{#2}}}
\nc{\ex}[1]{\mbox{e}^{\,\textstyle#1}}
\nc{\mat}[4]{\left(\begin{array}{cc}#1&#2\\#3&#4\end{array}\right)}
\nc{\som}[9]{\left(\begin{array}{ccc}#1&#2&#3\\#4&#5&#6\\#7&#8&#9%
\end{array}\right)}
\nc{\tr}{\mathop{\mbox{tr}}\nolimits}
\nc{\ad}{\mathop{\mbox{ad}}\nolimits}
\nc{\Tr}{\mathop{\mbox{Tr}}\nolimits}
\nc{\Det}{\mathop{\mbox{Det}}\nolimits}
\nc{\rk}{\mathop{\mbox{rk}}\nolimits}
\nc{\ra}{\rightarrow}
\nc{\Ra}{\Rightarrow}
\nc{\LRa}{\Leftrightarrow}
\nc{\ot}{\otimes}
\rnc{\ss}{\subset}
\nc{\nul}{\noindent\underline}
\nc{\non}{\nonumber\\}
\nc{\subs}[1]{{\vspace*{0.5cm}}%
{\noindent\underline{#1}}{\addcontentsline{toc}{subsection}{#1}}%
{\vspace*{0.3cm}}}
\nc{\zb}{\bar{z}}
\rnc{\lg}{\frak{g}}
\nc{\lt}{\frak{t}}
\nc{\lk}{\frak{k}}
\nc{\lh}{\frak{h}}
\nc{\pik}{\Pi_{\lk}}
\nc{\pip}{\Pi_{+}}
\nc{\pim}{\Pi_{-}}
\nc{\pih}{\Pi_{\lh}}
\nc{\jz}{J_{z}}
\nc{\jzh}{\jz^{\lh}}
\nc{\jzp}{\jz^{+}}
\nc{\jzm}{\jz^{-}}
\nc{\del}{\partial}
\nc{\dz}{\del_{z}}
\nc{\dzb}{\del_{\bar{z}}}
\nc{\az}{A_{z}}
\nc{\azb}{A_{\bar{z}}}
\nc{\g}{g^{-1}}
\nc{\dw}{\Delta_{W}}
\nc{\Ad}{{\mbox{Ad}}}
\nc{\ks}{Ka\-za\-ma-\-Su\-zu\-ki}
\nc{\KS}{\ks}
\nc{\ksm}{\ks\ model}
\rnc{\AA}{{\Bbb A}}
\nc{\BB}{{\Bbb B}}
\nc{\CC}{{\Bbb C}}
\nc{\PP}{{\Bbb P}}
\nc{\cpm}{\CC\PP(m)}
\nc{\cpn}{\CC\PP(n)}
\nc{\cp}[1]{\CC\PP(#1)}
\nc{\gmn}{G(m,m+n)}
\nc{\gmnk}{\gmn_{k}}
\nc{\cO}{{\cal O}}
\nc{\bcO}{\bar{\cO}}
\nc{\bO}{\bar{O}}
\nc{\oQ}{\overline{Q}}
\nc{\ie}{{\it i.e.~}}
\nc{\eg}{{\it e.g.~}}
\begin{document}
\global\parskip=4pt
\makeatother\begin{titlepage}
\begin{flushright}
{}
\end{flushright}
\vspace*{0.1in}
\begin{center}
{\Large\bf AdS/CFT correspondence and D1/D5 systems \\ 
in theories with 16 supercharges} \\ 
\vskip .3in
\makeatletter

\centerline{ Edi Gava\footnote{E-mail: gava@he.sissa.it}, \ 
Amine B. Hammou\footnote{E-mail: amine@sissa.it}, \ Jose F. Morales
\footnote{E-mail: morales@phys.uu.nl }, \
{\it and}
\ Kumar S. Narain \footnote{E-mail: narain@ictp.trieste.it} } 
\bigskip 
\centerline{\it SISSA, Trieste, Italy$^{1,2}$} 

\smallskip \centerline{\it The Abdus Salam ICTP, Trieste, Italy$^{1,4}$}

\smallskip \centerline{\it INFN, Sez. di Trieste,
Trieste, Italy$^{1,2}$} 

\smallskip \centerline{\it Spinoza Institute, Utrecht, The Netherlands $^3$} 

                                                                   
\end{center}

\vskip .10in
\begin{abstract}
\noindent 
We discuss spectra of $AdS_3$ supergravities, arising in the
near horizon geometry of D1/D5 systems in orbifolds/orientifolds
of type IIB theory with 16 supercharges. These include models studied
in a recent paper (hep-th/0012118), where the group action involves also
a shift along a transversal circle, as well as IIB/$\Omega I_4$, which
is dual to IIB on $K3$. After appropriate assignements of the orbifold 
group eigenvalues and degrees to the supergravity single 
particle spectrum, we compute the supergravity elliptic genus
and find agreement, in the expected regime of validity, with
the elliptic genus obtained using U-duality map from (4,4) CFTs
of U-dual backgrounds. Since this U-duality involves the exchange 
of KK momentum
$P$ and D1 charge $N$, it allows us to test
the (4,4) CFTs in the $P < N/4$ and $N < P/4$ regimes by two different 
supergravity duals.

\end{abstract}

\vfill


\makeatother
\end{titlepage}
\begin{small}

\end{small}

\setcounter{footnote}{0}

\tableofcontents

\section{Introduction}

One of the most interesting examples of
$AdS/CFT$ correspondence, proposed in \cite{maldacena},
relates type IIB string theory on 
$AdS_3\times S^3\times M$ to a CFT arising as the infrared
fixed point of the effective gauge theory governing a sysytem of  $Q_1$
D1-branes and $ Q_5$
D5-branes. The D5-branes wrap the internal manifold 
$M=T^4$ or $K3$ and are
parallel to the D1-branes, the common world-volume  
being identified with the boundary of $AdS_3$. 
Compared to the higher dimensional cases, testing the 
correspondence in this case is a more tractable problem, 
since the CFT's moduli space has points which correspond to exactly
solvable theories, given by the symmetric product CFTs
$M^N/S_N$ \cite{V,SV,W1,GKS,D1}.

Tests of this conjecture were  performed in
\cite{MS,DB1,DB2} (for $M=K3$) and \cite{mms} (for  $M=T^4$), 
where multiplicities of ground states of the D1/D5 ${\cal N}=(4,4)$ 
CFT were shown to agree with those of 
(chiral, chiral) primary states in the underlying supergravities. 
In supergravity the spectrum of 
(single particle) chiral primaries is determined
by group theory via Kaluza-Klein reduction,
while multiplicities in the boundary CFT  
can be read off from the index of the corresponding 
${\cal N}=(4,4)$ CFT. 
Although the supergravity
description is expected to be valid only for large values 
of $Q_1$ and $Q_5$, the correspondence was shown to 
work for all $N=Q_1Q_5$,                                                       
once a new additive quantum number, the {\it degree} $d$,
is introduced on the
supergravity side\cite{DB2}. This is a non-negative integer associated
to (chiral, chiral) primary states and
allows to cut-off multiparticle states and
implement the exclusion principle \cite{MS}: one
keeps only products of chiral primaries whose total
degree is $\leq N$.

Adopting the above prescription also for descendants
of chiral primaries. i.e. for states of the type
(chiral, anything), it was shown  in \cite{DB2} 
that the multiplicities obtained from supergravity 
agree, for states of low enough conformal weight $h\leq (N+1)/4$, 
with the multiplicities obtained  from the elliptic
genus of the boundary CFT.  

It is natural to explore the correspondence
between supergravity and boundary CFT
in other string theories with sixteen supercharges.
The aim is two-fold: on one hand this 
would provide additional examples of AdS/CFT correspondences 
in theories with 16 unbroken supercharges. 
On the other hand, and more importantly, one may hope to be able to learn 
more about the D1/D5 CFT from the dual supergravity description,
especially in cases, as in type I-like theories, 
where very little is known about the expected (4,0) CFTs.

In this paper we present a detailed analysis of the spectrum
of chiral primary states and their descendants in $CFT_2$/$AdS_3$
supergravity 
pairs, arising from the D1/D5 system in  
a class of freely acting $Z_2$ orientifolds of type IIB theory.
Correspondingly, the near horizon geometries are certain
freely acting $Z_2$ orbifolds of $AdS_3\times S^3\times T^4$. 
The associated boundary CFTs have been studied in 
great detail in \cite{ghmn}. In addition we 
will consider the $AdS_3$ supergravity corresponding to 
type $IIB/ \Omega I_4$. Although very little is known about the
CFT describing the boundary dynamics in this case, the relevant 
BPS counting formula can be derived, as we will see, via U-duality
from the  better understood D1D5 system in type IIB on $K3$.

The freely acting $Z_2$ group generators 
are defined by accompaning
the orbifold and/or orientifold actions $\Omega$, $I_4$, $\Omega I_4$    
with a shift $\sigma_{p_6}$ along a circle transverse to the
D1/D5 system,  with compact coordinate $X^6$. 
We refer to these theories as models $I$, $II$ and $III$
respectively. The analysis for D1/D5 systems in the presence of a
shift $\sigma_{p_1}$ longitudinal to 
the world-volume of the D1- and D5-branes,
was also performed in the $I_4$ case (model $IV$ in \cite{ghmn}),
but the AdS/CFT dictionary for this system becomes more involve and
we postpone a carefull study of it to elsewhere.  
Besides avoiding technical complications related to the gauge bundles
(twisted sectors), these models, with freely acting
orbifold group, actions are insteresting in their own right,
since D1/D5 states can be mapped via U-duality chains to 
fundamental string descriptions in terms of type IIB orbifolds generated by 
$(-)^{F_L} I_4 \sigma_{p_6}$, $(-)^{F_L}\sigma_{p_6}$ and $I_4\sigma_{p_6}$
respectively for the first three models (table 1.2 of \cite{ghmn})
and to an heterotic background 
with Wilson lines for the fourth one. 
 
In \cite{ghmn} the effective gauge theories were argued to flow
in the infrared to CFTs locally equivalent to the one 
appearing for the D1/D5 system in type IIB theory on 
$T^4\times S^1$, but with additional $Z_2$ global identifications
induced by the orbifold group actions \cite{ghmn}. The resulting
target spaces in the three models are of the form:
\be
{\cal M}_{\rm higgs}=\left( R^3\times S^1\times T^4
\times (T^4)^N/S_N \right)/Z_2
\label{mh}
\ee
The $Z_2$'s are generated by $(-)^{F_L}\, I_4^{\rm c.m.}$,
$  I_4^{\rm c.m.}\,I_4^{\rm sp} $ and $(-)^{F_L}\, I_4^{\rm sp}$ 
for the models $I$, $II$ and $III$ respectively, with  
$(-)^{F_L}$ the left moving spacetime fermionic number, 
$I_4^{\rm c.m.}$ the reflection of the first $T^4$ factor in
(\ref{mh}) and  $I_4^{\rm sp}$ the diagonal $Z_2$ reflection 
of the $N$ copies of $T^4$ in the symmetric product part. 
Pure D1/D5 systems correspond to states in the untwisted sectors
of (\ref{mh}). In a similar way, one can read off the BPS multiplicities
for the longitudinal shift case (model $IV$) from (\ref{mh}) with the
$Z_2$ induced by $I_4$, like in model $II$, but now with states
from the twisted sector.
The resulting CFTs are of type  ${\cal N}=(4,4)$  for  
models $II$, $IV$ and ${\cal N}=(4,0)$ for models $I$ and $III$, which
involve the $\Omega$ world sheet parity projection.

The spectrum of BPS charges and multiplicities of D1/D5 excitations
can be obtained from the elliptic genus
\bea
{\cal Z}{g_0\brack h_0}({\cal H}_N|q,\bar{q},y, \tilde{y})&=&
{\rm Tr}_{{\cal H}}\, q^{L_0-c/24}\,\bar{q}^{\bar{L}_0-c/24}
\,y^{J^3_0} \,\tilde{y}^{\bar{J}^3_0}\nonumber\\
&=&\sum
\,C{g_0\brack h_0}(h,\bar{h},\ell,\tilde{\ell})
\,q^{h}\,\bar{q}^{\bar{h}}\, y^\ell \, \tilde{y}^{\tilde{\ell}}
\label{zh}
\eea
evaluated in each of the CFT Hilbert spaces ${\cal H}_N$ defined by 
(\ref{mh}).  
The sum in (\ref{zh}) runs over $h,\bar{h},\ell,\tilde{\ell}$; 
$q=e^{2\pi i\tau}$ describe the genus-one 
worldsheet modulus, $\bar{L}_0$, $L_0$ are the Virasoro generators
and $\bar{J}_0^3, J_0^3$ are Cartan generators 
of an $SU(2)_R\times SU(2)_L$ 
current algebra to which the sources $y$ and 
$\tilde{y}$ couple respectively. $g_0$ and $h_0$ denote 
the boundary conditions for the various fields appearing in the
sigma model. 
In all the cases  results can be written as:
\bea
{\cal Z}(p,q,y,\tilde{y}) &=& \sum_{N} p^N  
Z({\cal H}_N|q,y,\tilde{y})\nonumber\\
&=&\sum_{h=0,{1\over 2}} Z_{cm}{0\brack h}(q,y,\tilde{y})
\, \hat{Z}_F{0\brack h}(p,y,\tilde{y}) 
\hat {Z}_{sym}{0\brack h}(p,q,y,\tilde{y})
\label{zpqf}
\eea
where $Z_{cm}{0\brack h}(q,y,\tilde{y})$ is the contribution
to the elliptic genus of the center of mass part
in (\ref{mh}) and $\hat{Z}_F{0\brack h}(p,y,\tilde{y}) 
\hat {Z}_{sym}{0\brack h}(p,q,y,\tilde{y})$ the contribution from
the symmetric product part. 
The latter can be computed using (4.17) in \cite{ghmn}, which
generalize the familiar symmetric product formulas 
derived in \cite{dmvv}.
We have isolated in (\ref{zpqf}) 
the ground state contribution 
$\hat{Z}_F{0\brack h}(p,y,\tilde{y})$ 
which admits always perturbative description in terms of U-dual 
fundamental strings carrying non trivial 
windings and momenta (see section 2 of \cite{ghmn} for details).
Indeed, one can easily see that the $q^0$ 
coefficient in the expansion of (\ref{zpqf}) 
\bea
&&Z{0\brack 0}(p,y,\tilde{y})
=y^2_- \tilde{y}^2_- \, \frac{\vartheta_1^2(y|p)
\vartheta_1^2(\tilde{y}|p)}
{\hat{\vartheta}_1(y \tilde{y}|p)
\hat{\vartheta}_1(y \tilde{y}^{-1}|p)\eta^6(p)}.
\label{iib}\\
&& Z_{I}{0 \brack {1\over 2}}(p,y,\tilde{y}) =
 {1\over 2}\,y_-^2\tilde{y}_+^2 \, {\vartheta_1^2(\tilde{y}|p)\over 
\hat{\vartheta}_1(y \tilde{y}|p)
\hat{\vartheta}_1(y \tilde{y}^{-1}|p)}
\,{\vartheta_2^2(y|p)\over \hat{\vartheta}_2^2(0|p)}\nonumber\\
&& Z_{II}{0 \brack {1\over 2}}
(p,y,\tilde{y})={1\over 2}\, y_-^2\tilde{y}_-^2 \,{ 
\vartheta_2^2(\tilde{y}|p)\over 
\hat{\vartheta}_1(y \tilde{y}|p)
\hat{\vartheta}_1(y \tilde{y}^{-1}|p)}
\,{\vartheta_2^2(y|p)\over \eta^6(p)}\nonumber\\
&& Z_{III}{0 \brack {1\over 2}}(p,y,\tilde{y}) =
 {1\over 2}\ y_+^2\tilde{y}_-^2 \,,{ \vartheta_2^2(\tilde{y}|p)\over 
\hat{\vartheta}_1(y \tilde{y}|p)
\hat{\vartheta}_1(y \tilde{y}^{-1}|p)}
\,{\vartheta_1^2(y|p)\over \hat{\vartheta}_2^2(0|p)}\label{d1d5t}\\
&& Z_{IV}{{1\over 2} \brack 0}(p,y,\tilde{y})= 16\,
y_-^2 \tilde{y}_-^2\, { 1\over
\hat{\vartheta}_1(y \tilde{y})
\hat{\vartheta}_1(y \tilde{y}^{-1})
\eta^{2}}{\eta^8\over \vartheta_4^8(0)}
\label{d1d5l}
\eea
reproduces, respectively, the multiplicities of fundamental strings
in type $IIB$ theory on $T^4\times S^1/g\sigma_{p_6}$ with
$g$ being $(-)^{F_L} I_4$, 
$(-)^{F_L}$ and $I_4$ for models $I$, $II$, $III$ and the ones
for heterotic strings moving on $T^5/Z_2^5$ for model $IV$
\cite{ghmn}.

These results support the CFTs proposals (\ref{mh}) at the two-charge
level but, as extensively discussed in \cite{ghmn}, the predictions
coming from (\ref{zpqf}) for the degeneracies of three charge 
states (D1-D5-KK) are not in agreement with basic constraints imposed
by U-duality. In particular, U-duality for $Q_5=1$, 
maps model $II$ to model $III$, while exchanging
the number of D1-branes $Q_1=N$ with the KK momentum.
This in turn  means that the elliptic genera of the two systems 
should be related by $p\leftrightarrow q$ exchange.
The problem arises because, even if one trusts just the
(4,4) CFT of  model $II$, the $p\leftrightarrow q$ 
exchange in its elliptic
genus does not give an expression admitting a CFT interpretation,  
assuming that in the CFT a 
$U(1)$ current algebra should be present. 
A similar problem
afflicts the more familiar $K3$ case: 
there, string-string duality 
requires that excitations at level $k$ of the CFT with $N=Q_1$
in $(K3)^N/S_N$
should be mapped to states 
at level $N$ in a CFT associated to 
$k=Q_1$ 
in type IIB/$\Omega I_4$. Again,
the elliptic genus (\ref{zpqf}) of  
$R^4\times (K3)^N/S_N$ does not have, after the $p\leftrightarrow q$ exchange,
the definite modular properties required by a CFT interpretation
\footnote{Notice that the CFT in question is the one that should
appear in the non supersymmetric sector of a (4,0) CFT.}.
In \cite{ghmn}, we proposed an explanation of this fact,
according to which U-duality maps the (4,4) CFTs above
to systems with vanishing RR 0- and 4-form backgrounds,
whereas the CFT we proposed for model $III$ is valid for non-trivial
RR backgrounds, which however cannot be connected continuously 
to the trivial background. Indeed, in model $III$ only
discrete ($Z_2$) values of RR 0- and 4- form are allowed, 
since these fields are odd under $\Omega$.      

Shedding some light on this issue, 
from the supergravity point of view,  
is the basic  motivation of the present work. The main result
will be that, in fact, the supergravity  
elliptic genus is, in the expected regime of
validity, in agreement
with the proposed CFT of model $II$, but,  
for both models $III$ and IIB/$\Omega I_4$,
it agrees  with the predictions of U-duality rather than 
with the  
counting formulas coming from the symmetric product 
CFT proposals in \cite{ghmn}. 
These results seem to support the arguments given in
the conclusions of \cite{ghmn} and recalled in the previous 
paragraph.
 
The paper is organized as follows: in section 2 we review 
the computation of the KK spectrum of ${\cal N}=(2,2)$ 
6-dimensional supergravity
on $AdS_3\times S^3$ and organize it in terms of short
multiplets of $SU(1,1|2)_R\times SU(1,1|2)_L$. The comparison
with boundary CFT is done for (chiral, chiral) states at finite
$N$. In section 3 we extend the analysis to our orbifold
models which involve (1,1) supergravities in 6-dimensions. 
We assign $Z_2$ eigenvalues and degree to the previously 
obtained short multiplets.  Using these assignments, in section 4, we  
compute the Poincare' polynomial and then the elliptic genus and compare
them with those of the CFT~s.  
In section 5, we perform a similar analysis for IIB/$\Omega I_4$,
the dual of IIB on $K3$.
Finally, in section 6, we give some conclusions.

\section{KK-reduction of $D=6$ ${\cal N}=(2,2)$ supergravity on 
$AdS_3\times S^3$}

In this section we will review the construction
of the supergravity multiparticle partition
function from the KK data for ${\cal N}=(2,2)$ 6-dimensional
supergravity, corresponding to type II theory compactified on 
$T^4$.  Our analysis will follow essentially \cite{DB1,DB2,mms}. 

The KK spectrum can be efficiently organized into
short supermultiplets of the $AdS_3$ supergroup 
$SU(1,1|2)_R\times SU(1,1|2)_L$.
This group is generated by the global generators $\{ L_{\pm 1}, L_0,
G^{\pm{1\over 2}}_{\pm{1\over 2}}, J_0^3, J_0^{\pm} \}_{L,R}$ of
the ${\cal N}=(4,4)$ Superconformal Algebra (SCA), 
whose bosonic part is identified with the
$SO(2,2)\times SO(4)$ isometry group of $AdS_3\times S^3$. We will 
denote by $({\bf m=2j})$ a short supermultiplet, which is a highest
weight representation  of the chiral SCA, 
with highest weight state given by the chiral primary state $h=j$. 
The whole representation
is constructed by acting on this state with the raising
operators  $\{ L_{-1}, G^{\pm{1\over 2}}_{-{1\over 2}}, J_0^{-} \}_{L,R}$.

The spectrum of KK harmonics on $S^3$ can be determined
essentially by group theory \cite{SS,dkss,DB1}.
To find  the $SO(4)$ (isometry group of the $S^3$)
representations that arise in the KK reduction,
one starts from the representations of the 
little Lorentz group $SO(4)$ under which  
the 6-dimensional massless states of
IIB supergravity, obtained by reducing the 10-dimensional 
theory on $T^4$, transform.
Identifying the $SO(3)$ tangent group of $S^3$ as
a subgroup of this $SO(4)$, one finds 
$SO(3)$ representations by decomposing the 
$SO(4)$ representations under $SO(3)$. 
Given a field which transforms as spin $j$ under 
the $SO(3)$ tangent group of $S^3$,
one then expands it using spin $j$ spherical harmonics. 
The latter in turn transform
under the isometry group of $S^3$ namely $SO(4) = SU(2) \times SU(2)$ as $(j_1,
j_2)$ where all $j_1$ and $j_2$ appear such that $(j_1 + j_2) \geq j \geq
|j_1-j_2|$.
 For the first few spin $j$ harmonics one
finds \footnote{In the following we will always label the $SU(2)$ spin-$j$
representations by specifying integers $2j$.}\\\\
\begin{tabular}{llll}
(2j) & KK-harmonics&  & \\
(0) & $(m,m)~~m=1,2,...$& & \\
(1) & $ (m,m\pm 1)~~m=1,2...$& & \\
(2) & $ (m,m)~~m=2,3,...$&$ (m,m\pm 2)~~m=1,2,..$ & \\
(3) & $ (m,m\pm 1)~~m=2,3...$&$ (m,m\pm 3)~~m=1,2...$&\\
(4) & $(m,m)~~m=3,4,...$ & $ (m,m\pm 2)~~m=2,3,$ &$ (m,m\pm 4)~~m=1,2$\\
\end{tabular}  \\\\
{\it Table 1: $S^3$ KK-harmonics}\\\\
For $m$ big enough, we  are then left with the following content 
of KK harmonics coming from six-dimensional massless states in the
representations of the $SO(4) = SU(2)\times SU(2)$ little group
\bea
&&(2,2)=(0)+(2)+(4)=3(m,m)+2(m,m\pm 2)+(m,m\pm 4)~~~{\rm Graviton}\nonumber\\ 
&&(2,1)=(1)+(3)=2(m,m\pm 1)+(m,m\pm 3)~~~~~~~~~~~{\rm 
Gravitino}\nonumber\\ 
&&(1,1)=(0)+(2)=2(m,m)+(m,m\pm 2)~~~~~~~~~~~~~~~~{\rm Vector}\nonumber\\
&&(2,0)=(2)=(m,m)+(m,m\pm 2)~~~~~~~~~~~~~~~~~~~~~~~~{\rm Self-dual~tensor}
\nonumber\\
&&(1,0)=(1)=(m,m\pm 1)~~~~~~~~~~~~~~~~~~~~~~~~~~~~~~~~~~~~
{\rm Left-spinor}\nonumber\\
&&(0,0)=(0)=(m,m)~~~~~~~~~~~~~~~~~~~~~~~~~~~~~~~~~~~~~~~~~{\rm Scalar}
\label{6d}
\eea
In particular using the field content of ${\cal N}=(2,2)$
supergravity in six dimensions (table 2 below) and 
organizing $(m,m')$ harmonics
into short supermultiplets of the ${\cal N}=(4,4)$ SCA defined by
\footnote{In the following the letters $m,m'$ will denote individual
components of the ${\cal N}=4$ supermultiplets while block letters 
${\bf m,m'}$ will denote
the entire supermultiplets labelled by their chiral primaries.}   
\be
({\bf m}, {\bf m'})=\sum^2_{i,j=0} \pmatrix{i \cr 2}\pmatrix{j \cr 2}
\, (m-i,m'-j),
\label{44}
\ee
one is left with the following spectrum of one-particle 
supergravity states \cite{DB1}
\begin{eqnarray}
{\cal H}_{\rm single~particle}&=&\oplus_{m\geq 1}
\left[({\bf m},{\bf m + 2})
+({\bf m+2}, {\bf m})+ 4 ({\bf m},{\bf m + 1})\right. 
\nonumber \\
&&\left. + 4({\bf m+1}, {\bf m})
+ 6 ({\bf m+1},{\bf m+1}) \right] +5({\bf 1},{\bf 1})
\label{short}
\end{eqnarray}

At this point, on the supergravity side there is no 
notion of finite $N$ physics. 
de Boer associated a new quantum number, the {\it degree},
 to the (chiral, chiral) 
primaries in order to truncate the supergravity 
states in a systematic way and 
reproduce the (chiral, chiral) states of 
the finite $N$ symmetric product boundary CFT. 
We recall the argument here since it will be 
useful for us later in order to
identify the $Z_2$ actions on the supergravity side. 
One starts from the 
Poincare' polynomial on the CFT side which captures 
the information about the chiral
primaries. Under spectral flow, chiral primaries in 
the NS sector go to the
ground states of the Ramond sector. 
The generating function for the Ramond ground states has
already been discussed in the last section.  
Flowing back from the Ramond states to the NS
states, using the relations (here we use the 
convention that a Ramond ground state
carries zero dimension)
\be
h_R=h_{NS}-j_{NS}\quad\quad  j_R=j_{NS}-{c \over 12}
\label{nsr}
\ee
the generating function of the  Poincare' polynomials 
for symmetric products of $T^4$ is given by
\be
P= \prod_{m=0}^{\infty}\prod_{r,s=0}^2 (1-p^{m+1} y^{m+r}
\tilde{y}^{m+s})^{-(-1)^{r+s} h_{r,s}}
\ee
where $h_{r,s}$ are the Hodge numbers of the torus. Explicitly 
$(-1)^{r+s} h_{r,s} = d(r).d(s)$ where $d(0) = d(2)=1$ and $d(1)=-2$.
The coefficient of $p^N$ in the expansion of $P$,
in the limit $N\rightarrow \infty$ is given by 
the residue of $P$ at $p=1$.
Indeed, $P$ has a first order pole, coming from the 
$m=r=s=0$ term, and the  residue, $P_{\infty}$, is then given by:
\be
P_{\infty}= \frac{(1-y)^2 (1-\tilde{y})^2}{(1-y^2)
(1-\tilde{y}^2)(1-y\tilde{y})^5} 
\prod_{m=1}^{\infty}\frac{(1-y^{m+1}\tilde{y}^m)^4 (1-y^m\tilde{y}^{m+1})^4}
{(1-y^{m+1}\tilde{y}^{m+1})^6 (1-y^{m+2}\tilde{y}^m)(1-y^m\tilde{y}^{m+2})}
\ee
This is exactly the content of the (chiral, chiral) 
states in the bulk supergravity
as seen from (\ref{short}), except for the states 
which involve either $j$ or $j'$
equal to zero. The missing states are precisely 
$(0,1), (1,0), (0,2)$ and $(2,0)$
states. It has been argued \cite{DB1,DB2} that 
these states do not correspond to propagating degrees of
freedom  in the bulk, but
neverthless appear at the boundary of $AdS_3$.
Including these states on the supergravity
side, one finds a complete matching of the 
chiral primaries in the supergravity and
the large $N$ limit of CFT. Note however that 
the $(0,0)$ state (corresponding to the identity operator) 
which is responsible for 
the simple pole in the CFT, does not appear in the
supergravity side. 

In order to extend this equality to finite $N$, 
de Boer introduced a notion of
degree. To each (chiral, chiral) primary  $(m,m')$, 
one asscociates a degree 
$d(m,m')$ which couples to the variable $p$ 
in such a way that the supergravity Poincare'
polynomial reproduces exactly the CFT one. 
The $(4,4)$ SCA is
assumed to commute with the degree and, as 
a result, all the descendants of a (c,c) 
primary carry the same degree as the primary itself. 
Thus the supergravity single
particle Hilbert space is described as
\be
{\cal H}_{\rm single~particle}=\oplus'_{m \geq 0} \, 
h_{r,s}\,({\bf m+r},{\bf
m+s})_{m+1}  
\label{shortpq}
\ee
where the subscript $m+1$ denotes the degree and 
the $\oplus'$ means $({\bf 0},{\bf 0})$ is 
omitted from the sum. In the above equation we have also added 
$({\bf 0},{\bf 1})$, $({\bf 1},{\bf 0})$, $({\bf 0},{\bf 2})$ and $({\bf
2},{\bf 0})$ supermultiplets to the supergravity single particle
states. While $({\bf 0},{\bf 2})$ and $({\bf 2},{\bf 0})$
supermultiplets contain the higher modes of the left and right ${\cal
N}=4$
super Virasoro generators, $({\bf 0},{\bf 1})$ and $({\bf 1},{\bf 0})$
contain the left and right $U(1)^4$ super Kac-Moody generators. 
The multiparticle Hilbert space is obtained as usual
by taking products of single particle 
states with the degrees being an additive quantum
number. The finite $N$ CFT Hilbert space is then conjectured to be
\be
{\cal H}^{CFT}_N = {\cal H}^{\rm multiparticle}|_{{\rm degree} \leq N}
\ee
By construction, the right-hand side reproduces the 
multiplicities of the (c,c) primaries of the 
CFT. Note that the $(1-p)^{-1}$ factor of 
the CFT appears in the gravity side
due to the fact that one allows all states with 
degree up to (and not just equal to) $N$.

\section{ Supergravity spectra for type
IIB orbifolds/orientifolds 
in the presence of transverse shifts}

In this section we will implement the  free $Z_2$ orbifold/orientifold
(models $I$, $II$, $III$ introduced above),  on the KK spectrum of 
6-dimensional supergravities on $AdS_3\times S^3$ obtained in the
previous section.
If we consider the radius $R_6$ of the circle, along which the shift 
is performed, very large, in such a way that the
space transverse to the D1/D5 system is effectively $R^4$,
then the near horizon geometry will still be
$AdS_3\times S^3 \times T^4$. However, in doing the KK reduction 
the various modes will come with non-trivial $Z_2$ phases due to
the orbifold group actions ($\Omega$, $I_4$
and $\Omega I_4$ according to the model) in the way we will specify below.   
  
The relevant $Z_2$-eigenvalues for 6-dimensional
massless fields of ${\cal N}=(2,2)$ supergravity, together with their
transformation properties under the little group $SO(4)$, 
are displayed in the following table:\\\\ 
\begin{tabular}{lllll}
$I$ & $II$ & $III$ & \\  
$\Omega$& $I_4$ &$\Omega I_4$& $bosons$ & $fermions$\\
$+$ &$ +$ &$+$& (2,2)+(0,2)+(2,0)+17(0,0) & 2(1,2) + 10(1,0) \\
$-$&$ +$&$-$& 4(0,2)+4(2,0)+8(0,0) & 2(1,2)+ 10(1,0)\\
$+$&$ -$&$-$& 8(1,1) & 2(2,1)+ 10(0,1) \\
$-$&$ -$&$+$& 8(1,1)& 2(2,1) + 10(0,1)\\
\end{tabular}\\\\\
{\it Table 2: SO(4) field content with $Z_2$ eigenvalues}\\\\
For example, the first row in table 2 is the contribution of
the 6-dimensional metric $G_{\mu\nu}$, the RR two-form $B^{R}_{\mu\nu}$,
the dilaton 
and the scalars associated to the internal components 
$G_{ij}$, $B_{ij}^{RR}$ of the metric and RR two-form. 
They are clearly even under all three orbifold group actions.
One can similarly obtain the other contributions with the corresponding
eigenvalues. As expected, one sees from the 
above table that the model $II$ has
(2,0) supersymmetry while models $I$ and  $III$ 
have (1,1) supersymmetry in 6-dimensions.

One now proceeds as in the $T^4$ case, decomposes 
these fields into representations
of the $SO(3)$ tangent group of $S^3$ 
(which is the diagonal subgroup of the $SU(2)
\times SU(2)$ little group). 
At this point the information of the chirality is lost 
and one finds that all the three models have the 
same representation content of
$SO(3)$ and, as a result, will have the same $S^3$ 
spherical harmonics. One would then
naively conclude that all the three models 
are identical on the supergravity side.
The subtle point however is that for a given $S^3$
spherical harmonic, the dimensions $L_0$ are different 
for the three models. To compute the dimensions one would
have to solve the equations of motion for various fields. However,
given the fact that all the systems  at least have (4,0) supersymmetry,
which fixes the $\bar{L}_0$ eigenvalues in terms of the data encoded
in the spherical harmonics, we can deduce the $L_0$ eigenvalues
by determining the spin $s = L_0 - \bar{L}_0$, as has been done 
in \cite{DB1}. But before doing so, we give an intuitive
argument to determine the $Z_2$ actions on the different representations.

Let us consider the $SO(3)$ spin 2 and 3/2
states. Spin 2 appears only from the graviton 
(i.e. from the first row), while
spin 3/2 appears from gravitini, and therefore 
2 from each row in table 2. The spherical
harmonics of spin 2 and spin 3/2, among others, 
would contain $(m,m+4)$ and $(m,m+3)$
states respectively. Now $(m,m+4)$ appears in 
the short multiplet $({\bf m}, {\bf
m+2})$. This short multiplet also contains 
4 states of the type $(m,m+3)$, two of
which appear by applying left supersymmetry 
generator $G^i_{-\frac{1}{2}}$ once and
the other two by applying two left supersymetry generators and one right
supersymmetry generator $G^1_{-\frac{1}{2}}
G^2_{-\frac{1}{2}}\tilde{G}^i_{-\frac{1}{2}}$. The remaining 4 $(m,m+3)$ states
appear in the 4 short multiplets $({\bf m}, {\bf m+1})$ by applying 2 left
supersymmetry generators. The 4 $(m,m+3)$ states 
that appear in the short multiplet
$({\bf m}, {\bf m+2})$ must come from the 
first two rows in the above table. This
is because in model $II$ the $Z_2$ action must 
commute with the (4,4) supersymmetry
and it is the first two rows that come 
with positive $Z_2$ eigenvalue. The
remaining 4 $(m,m+3)$ states appearing in the 4 short multiplets 
$({\bf m}, {\bf m+1})$ must therefore come 
from the last two rows. In particular,
it also implies that in model $II$, 
the short multiplets $({\bf m}, {\bf
m+2})$ must appear with $+$ eigenvalues and $({\bf m}, {\bf
m+1})$ must appear with $-$ eigenvalue. Having identified the 4 $(m,m+3)$ 
states 
appearing in the short multiplet $({\bf m}, {\bf m+2})$ 
as the ones coming from the
first two rows, it now becomes clear that in 
models $I$ and $III$ the $Z_2$ actions
would not commute with (4,4) supersymmetry. 
In fact, they only commute with (4,0)
supersymmetry, with the left supersymmetry 
generator $G^i_{-\frac{1}{2}}$ having
negative eigenvalues. This
argument also shows that of the 4 
$({\bf m}, {\bf m+1})$ short multiplets that
appear from the last two rows, the (c,c) 
primaries of the two should come with $+$
eigenvalue and two with $-$, the role of $+$ and $-$ 
being exchanged between models $I$
and $III$. Furthermore we conclude that 
while $\tilde{G}^i_{-\frac{1}{2}}$ moves 
the states within each row, $G^i_{-\frac{1}{2}}$ moves vertically between
first two rows as well as the last two rows.

We can repeat the above procedure for all the states and  using the fact that 
the $Z_2$ of model $II$ commutes with (4,4) supersymmetry, show that all the 
short multiplets of the type $({\bf m+r}, {\bf m+s})$ 
with $r+s$ even come from the
first two rows and all the ones with $r+s$ odd come from the last two rows. 
Furthermore, the (c,c) primaries of these short multiplets 
appear in the following
way: $(m,m+2)$, $(m+2,m)$ and two $(m,m)$ 
appear in the first row, 4 $(m,m)$ appear
in the second row, and in the third and fourth 
row each 2 $(m,m+1)$ and 2 $(m+1,m)$.

We will now justify the above result by explicitly computing the $L_0$
eigenvalues. We will decompose the 6-dimensional (2,0) or (1,1) 
supermultiplets into (1,0) supermultiplets. The latters, upon KK
reduction
on $AdS_3 \times S^3$, give rise to (4,0) supermultiplets of 
$SU(1,1|2)_{R}\times SU(2)_{L}\times SU(1,1)_{L}$, which can be
labelled by $({\bf m}, m';h )$ where ${\bf m}$ denotes the $2J_3$
eigenvalue of the chiral primary  of the $SU(1,1|2)_R$ superalgebra,
$m'/2$ and $h$  denote the isospin and $L_0$ eigenvalue under the
left 
$SU(2)_{L} \times SU(1,1)_{L}$ subgroup of the
isometry of $S^3 \times AdS_3$. Even though there is no supersymmetry
in the left sector, $h$ is obtained  by determining the helicity
as in
\cite{DB1}. The method is as follows. Given a field which transforms
as $(n_1, n_2)$ under the little group $SU(2)\times SU(2)$ (which is
not the same as the isometry of $S^3$), its $S^3$ harmonic labelled by
$(m,m')$ under the isometry group carries, a helicity $s$ which is
given by all possible values of $y_1-y_2$ subject to the condition 
$y_1+y_2= m'-m$, with $y_1$ and $y_2$ being
the $J_3$ component of the $SU(2)$ in the representations $(n_1)$ and
$(n_2)$ respectively. 

This analysis has been done in \cite{DB1} for
graviton, vector, tensor and hypermultiplets. We label these
multiplets as ${\bf G}$, ${\bf T}$, ${\bf V}$ and ${\bf H}$ respectively. 
The result is
\begin{eqnarray}
{\bf G}&:&~~~~~ ({\bf m}, m+2; \frac{m+2}{2})+({\bf m}, m;
\frac{m}{2})
+({\bf m}, m; \frac{m+4}{2})\nonumber \\
&~&~~~~~+({\bf m}, m-2; \frac{m+2}{2})+
({\bf m}, m-2; \frac{m-2}{2})+({\bf m}, m-2; \frac{m}{2}) \nonumber\\
{\bf T}&:&~~~~~ ({\bf m}, m; \frac{m}{2})+({\bf m}, m-2; \frac{m+2}{2})
\nonumber \\
{\bf V}&:&~~~~~ ({\bf m}, m; \frac{m+2}{2}) +({\bf m}, m-2; \frac{m}{2})
\nonumber \\
{\bf H}&:&~~~~~ 2({\bf m}, m-1; \frac{m+1}{2})
\label{GTVH}
\end{eqnarray}
Note that the tensor and vector multiplets, although they come with same
spherical harmonics, carry different $L_0$ eigenvalues. Besides these
standard
(1,0) multiplets, we also need two multiplets with highest spin 3/2
which
appear in the decomposition of (2,0) and (1,1) representations in
terms
of (1,0) representation. These multiplets (say ${\bf U}$ and ${\bf 
\tilde{U}}$ ) contain
the following representations under the little group:
\be
{\bf U}= (1,2) + 2(0,2),  ~~~~~~~~~~~ 
{\bf \tilde{U}}= (2,1) + 2(1,1) +(0,1)
\ee
KK analysis for these two multiplets give:
\begin{eqnarray}
{\bf U}&:&~~~~~({\bf m}, m-3; \frac{m-1}{2})+({\bf m}, m-1; 
\frac{m+1}{2})+
({\bf m}, m+1; \frac{m+3}{2})\nonumber\\
{\bf \tilde{U}}&:&~~~~~({\bf m}, m-3; \frac{m+1}{2})+({\bf m}, m-1; 
\frac{m+3}{2})+
({\bf m}, m-1; \frac{m-1}{2})\nonumber\\
&~&~~~~~+({\bf m}, m+1; \frac{m+1}{2})
\label{UW}
\end{eqnarray}
To illustrate how we obtained the above multiplets let us consider
${\bf U}$. The (1,2) and (0,2) give rise to following components 
\begin{eqnarray}
(1,2)&:&~~~~ (m-2,m+1; h-\frac{1}{2},h) + (m,m+1;
h-\frac{3}{2},h)\nonumber\\&~&~~~~ + (m-2,m-1; h+\frac{1}{2},h) 
+(m,m-1; h+\frac{1}{2} \pm 1,h) \nonumber \\
&~&~~~~
+(m-2,m-3; h+\frac{3}{2},h) +
(m,m-3; h+\frac{1}{2},h) 
\nonumber\\
2(0,2)&:&~~~~ 2(m-1,m+1; h-1,h) +2(m-1,m-1; h,h)
\nonumber \\
&~&~~~~ +2(m-1,m-3; h+1,h)
\end{eqnarray}
Here the first and second entries are twice the spin under the  right
and  left  $SU(2)$s 
and the third and fourth entries are the $\bar{L}_0$ and $L_0$
eigenvalues. $h$ at this stage is undetermined but $\bar{h}$ is given
in terms of $h$ plus the helicity. The fact that these states must
organize in at least (4,0)
supermultiplets, fixes the values of $h$ and one gets  the result
(\ref{UW}). It is interesting to note that even in the
non-supersymmetric sector $h$ satisfies the bound $h \geq m'/2$ !!

Using the fact that the (2,0) gravitational and tensor multiplets 
are given by 
\be
{\bf G_{(2,0)}} = {\bf G} + 2 {\bf U}, ~~~~~~~~~~ 
{\bf T_{(2,0)}}= {\bf T} + {\bf H}
\ee
we find that the dimensions $h$ are such that the (4,0)
representations
combine to form (4,4) representations as expected. The result is
\begin{eqnarray}
{\bf G_{(2,0)}} &=& ({\bf m},{\bf m+2})+  ({\bf m},{\bf m})+
 ({\bf m},{\bf m-2})
\nonumber\\
{\bf T_{(2,0)}} &=&  ({\bf m},{\bf m})
\end{eqnarray}

On the other hand the (1,1) gravitational and vector multiplets are
\be
{\bf G_{(1,1)}} = {\bf G} + 2 {\bf \tilde{U}} + {\bf T}, 
~~~~~~~~~~~~ 
{\bf V_{(1,1)}} = {\bf V} +{\bf H}
\label{gv11}
\ee
It is easy to see from eqs.(\ref{GTVH},\ref{UW}) that the dimensions $h$ 
are such
that the right hand side of (\ref{gv11}) cannot be organized as
(4,4)
multiplets. Indeed in ${\bf V_{(1,1)}}$ there is no state with $h=m'/2$. 
In
fact for all the states in ${\bf V_{(1,1)}}$, $h=(m'+1)/2$. Therefore
formally
we can express the (1,1) vector multiplet as the first left
descendents
of (4,4) multiplets:
\be
2 {\bf V_{(1,1)}} =  ({\bf m},{\bf m+1})^- + 2 ({\bf m},{\bf m})^- +
 ({\bf m},{\bf m-1})^- 
\ee
where the superscript $-$ indicates that we should keep only the states 
with odd number of $G_{-1/2}$ acting on (c,c) primary. Similarly
\begin{eqnarray}
{\bf G_{(1,1)}} &=&  ({\bf m},{\bf m+2})^+ +
 2({\bf m},{\bf m+1})^+ +  2({\bf m},{\bf m})^+ \nonumber\\
&~& + 2({\bf m},{\bf m-1})^+ +  ({\bf m},{\bf m-2})^+
\end{eqnarray}
where the superscript $+$ indicates that we should keep only states with
even number of $G_{-1/2}$ acting on the (c,c) primary.
Note that the superscript $\pm$ are in fact the $Z_2$ assignments to the
various (4,4) multiplets appearing in the parent IIB theory on $T^4$,
and 
the correlation of $\pm$ with even and odd $G_{-1/2}$ indicates that 
$Z_2$ anticommutes with $G_{-1/2}$. This is exactly the result we
obtained from the earlier intuitive argument. The fact that in (1,1) 
supergravity the states cannot be organized in terms of complete
(4,4) multiplets is in agreement with string theory description of 
IIA on $AdS_3 \times S^3 \times K3$ \cite{GKS}, where it is shown that
only (4,0) supersymmetry survives.

Accepting the above composition of the short 
multiplets, we still have to assign 
degree to them. This has already been done for the 
$T^4$ as in eq.({\ref{shortpq}), but
here we need more refinement. We need to assign 
degrees for (c,c) primaries coming 
from each row separately. For example, there are 
altogether 6 (c,c) primaries of the
form $ (m, m)$, 4 coming from the second row 
while 2 from the first row of table 2. Given the
fact that they should be organized as  
1 $(m,m)$, 4 $(m+1,m+1)$ and 1 $(m+2,m+2)$
having all degree $m+1$, with $m \geq 0$, we 
note there should be only 5 $ (1,1)$
primaries. In the first row, indeed, 
there is one less $(1,1)$ compared to general
$(m,m)$s. This allows us to deduce that 
the $(m+2,m+2)$ series comes from the first
row. Since $(m,m)$ series corresponds to 
$h_{0,0}$ one would expect that for all
the three models this should be in the spectrum, 
therefore it is natural to assume
that it too comes from the first row. Finally, 
we are left with 4 $(m+1,m+1)$, which
should all be in the second row. 
A more delicate question is regarding 4 $(m,m+1)$
and 4$(m+1,m)$ type (c,c) primaries. From the $T^4$ analysis we know 
that there
should be 2 $(m,m+1)$ and 2 $(m+1, m+2)$ each with degree 
$m+1$ which should be
distributed between third and fourth rows. 
We do not know of any a priori reason 
to choose a particular assignment, but again anticipating the 
agreement with the
U-dual fundamental theory, we assign  2 $(m,m+1)$ in the third 
row and 2
$(m+1,m+2)$ in the fourth row. Similarly, we 
assign 2 $(m+2,m+1)$ in the third row
and the 2 $(m+1,m)$ in the fouth row, each with degree $m+1$. 

To conclude, we have the following $Z_2$ assignments for various models. 
\be
{\cal H}^A_{\rm single~particle}=\oplus'_{m \geq 0} \, 
h_{r,s}\,({\bf m+r},{\bf
m+s})^{\epsilon_A(r,s)}_{m+1}  
\label{shortrs}
\ee
where $A=I,II,III$, the subscript $m+1$ denotes the 
degree  and $\epsilon_A(r,s)$
are the $Z_2$ eigenvalues of the
multiplets and are given by:
\be
\epsilon^{I}(r,s) = (-1)^s,  ~~~~~~~~ \epsilon^{II}(r,s) =(-1)^{r+s}, 
~~~~~~~~\epsilon^{III}(r,s) = (-1)^r
\label{phases}
\ee
For models $I$ and $III$, although we have 
grouped the states in terms of 
the original (4,4) multiplets, we have to remember that 
$G^i_{-\frac{1}{2}}$ anticommutes
with the $Z_2$'s, and therefore the descendants that involve odd numbers of
$G^i_{-\frac{1}{2}}$'s will appear with an extra minus sign under the 
$Z_2$ action.

Let us now try to understand the $Z_2$ actions
defined in eqs.(\ref{shortrs}), (\ref{phases}) as geometric actions on $T^4$. 
The degree 1
i.e $m=0$ corresponds to a single copy of $T^4$. The (c,c)
primaries are generated by the two chiral left moving and right moving
fermions  $\psi_i$ and $\tilde{\psi}_i$ ($i=1,2$). The $h_{r,s}$
numbers of $(r,s)$ (c,c)
primaries
are then simply $\tilde{\psi}^r \psi^s$ where we have suppressed the
subscripts
$i,j$ on the fermions. The $Z_2$ actions in (\ref{shortrs}), (\ref{phases}) 
imply that $g_I$
reflects $\psi$, $g_{II}$ reflects both $\psi$ and
$\tilde{\psi}$
and $g_{III}$ reflects $\tilde{\psi}$. One can now deduce the $Z_2$ 
action on
the bosonic fields $\partial X$ and $\bar{\partial} X$, which are the
superpartners of $\psi$ and $\tilde{\psi}$ respectively, from its
action on supersymmetry generators $G$ and $\tilde{G}$. The result can
be summarized as $g_I = (-1)^{F_L}$, $g_{II} = I_4$ and $g_{III}=
(-1)^{F_L}.I_4$.  These are indeed the  $Z_2$ actions proposed in
\cite{ghmn} for the three symmetric product CFT~s.

Before proceeding further, we would like to comment on a puzzle.
Note that for model $I$ and $III$, the graviton and vector multiplets
have to  be assigned
the degrees in different ways. In particular for model $I$
\begin{eqnarray}
{\bf V_{(1,1)}} &=& \sum_{m=0}^{\infty}\sum_{r=0}^2  |d(r)|({\bf m+r}, 
{\bf m+s})^-_{m+1} ;~~~~~~~~~~ s=1
\nonumber\\
{\bf G_{(1,1)}} = &=& \sum_{m=0}^{\infty}\sum_{s=0,2}\sum_{r=0}^2  
|d(r)|({\bf m+r}, {\bf m+s})^+_{m+1}
\label{vI}
\end{eqnarray}
with similar expressions for model $III$ but with the role of "$r$" and
"$s$" exchanged. Although the above choice is required to obtain 
the correct Poincare' polynomials as dictated by U-duality,
it is a bit puzzling why these two models, which are both 6-dimensional
(1,1) supergravity theories, give rise to different results. However,
one can see that the difference lies in the following two facts:

1) the definition of degree,
whose understanding itself is outside the realm of
supergravity. Indeed
in each of these models, for $m \geq 1$, two of the (c,c) primaries 
$(m+1,m)$ 
come with positive
eigenvalue and the remaining two with negative eigenvalue under the
$Z_2$ action. For model $I$, we assign degree $m+2$ to the two
positive eigenstates and $m+1$ to the negative eigenstates, and 
vice versa for
model $III$. Similarly, for (c,c) primaries $(m,m+1)$ for $m \geq 1$
there are again two positive and two negative eigenstates. For model
$I$, the degree of the positive eigenstates is $m+1$ and that of the
negative
eigenstates is $m+2$, and vice versa for model $III$. We do not have
any a priori reason for this assignment, however the  
relation between the three $Z_2$~s namely $g_I. g_{II} = g_{III}$
fixes the degree for model $III$ once one
assumes it for model $I$.

2) the two $({\bf 0,1})$ and two $({\bf 1,0})$ are assigned
eigenvalues
minus and plus respectively for model $I$ and vice versa for model
$III$. These multiplets contain the left and right $U(1)^4$ super
Kac-moody generators respectively. They in fact do not represent 
propagating degrees
of freedom in the bulk and therefore they are not contained in the
supergravity spectrum. They should appear as large gauge
transformations that do not vanish at the boundary of $AdS_3$, in 
the same way as the $({\bf 2,0})$ and $({\bf 0,2})$ multiplets arise
from large super-diffeomorphisms. In the 6-dimensional theory obtained by
compactifying IIB on $T^4$, there are 16 $U(1)$ gauge fields with 4
each coming from the reduction of the metric, NS-NS anti-symmetric tensor,
and RR 2- and 4-forms. Of these only 8 gauge transformations
should be non-vanishing at the $AdS_3$ boundary, giving rise to
$U(1)^4_L
\times U(1)^4_R$ current algebra. 
The $Z_2$ actions we have defined here indicate that these 8 gauge
fields are the ones coming from the $T^4$ reduction of the metric and
the RR 2-form. Indeed we have assigned 2 $(1,0)$ and 2 $(0,1)$ chiral 
primaries in the third and the fourth rows of table 2
respectively. The $U(1)$ currents are the first descendents of the
chiral primary. Using now the fact that $\tilde{G}_{-1/2}$ and
$G_{-1/2}$ move the states horizontally and vertically, respectively, 
between the
third and the fourth rows, it is easy to see that all the $U(1)$
currents appear in the third row. The 6-dimensional vectors which
appear in the third row carry positive eigenvalues under $\Omega$
and therefore must come from the metric and the RR 2-form.

One may wonder if this could be understood directly
from a supergravity analysis. However 4 T-dualities on $T^4$ exchanges
model $I$ and $III$, so it would appear that if in model $I$, the
metric and RR 2-forms give rise to the $U(1)$ currents, then in model 
$III$ the NS-NS 2-form and RR 4-form should generate these currents. 
This would be in contradiction with our proposal here. The point is,
that under the T-duality, $Q_1$ and $Q_5$ get exchanged. Therefore, 
more precisely our proposal is, that for $Q_1 >> Q_5$, the $U(1)$
currents come from the metric and RR 2-form. Supergravity analysis
on the other hand depends only on the product $Q_1.Q_5$ and not the ratio
$Q_1/Q_5$ therefore it seems unlikely that the answer can be found
there. The world sheet approach \cite{GKS} for fundamental string moving
in  $AdS_3\times S^3
\times T^4$, corresponding to the near horizon geometry of $Q_5$ NS5 branes 
and $Q_1$ fundamental strings, depends on the individual values of
$Q_1$ and $Q_5$. In this framework the weakly coupled string 
(for a fixed order one volume of $T^4$) corresponds to precisely the regime
$Q_1 >> Q_5$. The $U(1)$ currents on the $AdS_3$ boundary come from the
NS-NS sector of the string theory and therefore should appear from the 
bulk gauge fields arising from metric and NS-NS 2-forms \footnote{In
fact an attempt to find the $U(1)$ currents corresponding to the RR gauge
fields has not succeeded so far \cite{KS}.}. This is
exactly the S-dual version of our proposal here.  
 
\section{Comparison between supergravity and CFT elliptic genera}

Now we can construct the multiparticle Hilbert space 
${\cal H}_{\rm multiparticle}$ as usual and
identify the finite $N$ CFT Hilbert space with 
the subset of states in ${\cal H}_{\rm multiparticle}$ 
that have degree less than or equal to $N$. The generating function of 
the Poincare' polynomials, in the sector which is projected by $Z_2$, is
\be
\frac{1}{1-p} {\rm Tr}_{(c,c) \in {\cal H}_{\rm multiparticle}} g^A p^{m+1} 
y^{\ell}
\tilde{y}^{\bar{\ell}}
\label{poincmult}
\ee
where $m+1$, $\ell$ and $\bar{\ell}$ are, respectively, 
the degree, and the left- and
right-moving values of twice the $SU(2)$ spins. 
$g^A$ are the generators of the various
$Z_2$'s. The prefactor $\frac{1}{1-p}$ just takes 
into account the fact that finite
$N$ CFT Hilbert space is identified with states in ${\cal H}_{\rm 
multiparticle}$  that have degree up to $N$. 
One can carry out the trace over all
(c,c) primaries above with the result 
\be
\prod_{m=0}^{\infty}\prod_{r,s=0}^2
(1-\epsilon^A(r,s)p^{m+1}y^{m+r}\tilde{y}^{m+s})^{-(-)^{r+s}h_{r,s}}
\ee
It is easy to see that these exactly reproduce the CFT results 
(\ref{d1d5t}) for the three models under 
spectral flow from Ramond to NS sector. Note that although  models $I$ and
$III$ we have only (4,0) supersymmetry, they arise as untwisted sector
of orbifold of (4,4) theory and we can use the spectral flow in the
parent
(4,4) theory. More generally we can still define a notion of spectral
flow using the $SU(2)$
level $n$
current algebra in the non-supersymmetric sector. By spectral flow
from ``Ramond'' to ``NS'' in the non-supersymmetric sector we then mean the
map
from spin-$j$ to spin-$(\frac{n}{2}-j)$ characters. The unitarity of the 
``Ramond'' sector CFT then implies the bound $h_{NS}
 \geq j_{NS}/2$ in the ``NS'' sector. The states contributing to
the Poincare' polynomials are the ones which saturate this bound and
therefore
correspond to the ground states in the ``Ramond'' sector.

One can now, following de Boer, check the 
correspondence beyond the (c,c) primaries.
The idea is to construct the finite $N$ elliptic 
genus which is obtained by taking 
the trace over states of the form chiral on the right-moving 
sector and any state on the
left-moving sector, and setting $\tilde{y} = \bar{q}^{-1/2}$. 
The comparison of the states can, of
course, only be made for dimensions much 
less than $N$, since otherwise gravity
approximation would break down. In \cite{DB1} de Boer showed that 
for the K3 case the matching
of the states goes all the way upto left dimension 
equal to $(N+1)/4$. This is
exactly the bound at which black hole is expected to form. 
In the right-moving sector
arbitrary chiral states are allowed, and they 
have a bound on the dimension 
which is of order $N/2$. Setting  
$\tilde{y} = \bar{q}^{-1/2}$ for right-chiral states, one actually 
looses the information about which chiral state 
is being traced over. One might ask
a more refined question of matching the states 
between CFT and supergravity for a fixed 
right chiral state. One can show that this does 
not work even in the $K3$ case. The
point is that there could be states of the form $(m,m')$, 2 $(m,m'-1)$ and
$(m,m'-2)$, where $m$ denotes an arbitrary left-moving state, 
while the states appearing on
the right-moving sector $m'$, $m'-1$ and $m'-2$ are 
all chiral states. Setting $\tilde{y} = 
\bar{q}^{-1/2}$ the contribution of these 4 states to the supertrace vanishes.
Therefore the fact that the states on the two sides do not match for arbitrary
$\tilde{y}$ implies that discrepancy comes in the combinations of the above 4
states. On the other hand, these 4 states can 
in principle combine to form a long
multiplet on the right-moving sector. So, it appears 
that as one changes the parameters 
of the theory from a region where CFT is valid to the region where gravity
approximation is valid, some of such combinations of 
chiral states become non-chiral
and form long multiplets. It would be interesting to understand the precise
mechanism of how this happens. However, in the 
following we will set $\tilde{y} = 
\bar{q}^{-1/2}$ and compare the elliptic genus on the two sides.

In model $I$, unfortunately the elliptic genus is zero, 
but here one can repeat the
analysis of \cite{mms} and take 2 derivatives with respect to $\tilde{y}$ 
before setting it to $\bar{q}^{-1/2}$. In this case however we do not get any
information; in fact for states satisfying the bound $h \leq (N+1)/4$ only
the ground state contributes as shown in \cite{mms}. 

Models $II$ and $III$ are more interesting, 
since for $\tilde{y}=\bar{q}^{-1/2}$
the elliptic genus does not vanish. The elliptic genus for CFT has 
already been 
obtained in section 4 of \cite{ghmn}, where the 
appropriate $Z_2$ actions
for models $II$ and $III$ have been used. 
The elliptic genus for supergravity is computed as follows.
We start from the supertrace over the states in the single particle Hilbert
space which are of the form chiral in the right-moving 
sector and arbitrary state from the
left-moving one. There will be two sectors, 
one where the identity element of $Z_2$ is inserted
and the second where the generator $g^A$ is inserted. 
\begin{eqnarray}
Z(p,q,y) &\equiv& \frac{1}{2}{\rm tr}  p^n q^m y^{\ell} \equiv
\sum_{n,m,\ell}\frac{1}{2}[c^{A+}_{\rm sgr}(n,m,\ell)+c^{A-}_{\rm sgr}
(n,m,\ell)] 
p^n q^m y^{\ell} 
\nonumber\\
Z^A(p,q,y) &\equiv& \frac{1}{2}{\rm tr} g^A p^n q^m y^{\ell} \equiv
\sum_{n,m,\ell}\frac{1}{2}[c^{A+}_{\rm sgr}(n,m,\ell)-c^{A-}_{\rm sgr}(n,m,\ell)] 
p^n q^m y^{\ell} 
\label{egsgrsp}
\end{eqnarray}
where $n$, $m$ and $\ell$ are the degree, left dimension and twice the left $J_3$
quantum numbers respectively.
The elliptic genus for the multi-particle states is then
\be
Z_{\rm multiparticle}^A(p,q,y) = \frac{1}{1-p}\prod_{n,m,\ell}(1-p^n q^m
y^{\ell})^{-c^{A+}_{\rm sgr}(n,m,\ell)}.(1+p^n q^m 
y^{\ell})^{-c^{A-}_{\rm sgr}(n,m,\ell)} 
\ee
where again $(1-p)^{-1}$ is included for the same reason as in 
(\ref{poincmult}).  
This factor can actually be absorbed inside the product by
redefining $c^{A+}_{\rm sgr}(n,m,\ell) \rightarrow c^{A+}_{\rm sgr}
(n,m,\ell) + \frac{1}{2}\delta_{n1}\delta_{m0}\delta_{\ell 0}$ which
is equivalent to $Z \rightarrow Z+\frac{p}{2}$ and $Z^A \rightarrow
Z^A+\frac{p}{2}$. In the following $Z$, $Z^A$ and $c$'s will always
denote these redefined quantities. 

$Z(p,q,y)$ vanishes for
$\tilde{y} = \bar{q}^{-1/2}$ since the elliptic genus for $T^4$ case
is 1. The fact that the trace
with identity insertion vanishes implies $c^{A+}_{\rm sgr}(n,m,\ell) =
-c^{A-}_{\rm sgr}(n,m,\ell)$
The elliptic genus for the multi-particle states is then
\be
Z_{\rm multiparticle}^A(p,q,y) 
=\prod_{n,m,\ell}\left[ \frac{1+p^n q^m y^{\ell}}{1-p^n q^m
y^{\ell}}
\right]^{c^{A+}_{\rm sgr}(n,m,\ell)}.
\label{egsgrmp}
\ee

To compute the $c$'s we write explicitly the trace in (\ref{egsgrsp}) as
\bea
Z^A(p,q,y) =&&\frac{1}{2} \sum_{m,r,s}
\sum_{t=0}^{min(2,m+s)} {\sum_{k=0}^{\infty}}' d(r)d(s)d^A(t)\epsilon^A(r,s)
p^{m+1}q^{\frac{m+s+t}{2}+k}\nonumber\\ &&\times \sum_{j=0}^{m+s-t}y^{m+s-t-2j}    
\label{egsgrsp1}
\eea
where we have used the fact that $(-1)^{r+s} h_{r,s} = 
d(r).d(s)$ with $d(0)=d(2)=1$
and $d(1)=-2$. The sum over $m,r,s$ takes into account all (c,c) primaries 
$(m+r,m+s)$. We have also included here the $m=r=s=0$, which is not a 
(c,c) primary in supergravity, to take into 
account the redefined $Z$'s including a shift by $p/2$.  
The sum over $k$ takes into account
the descendents coming from
applying $L_{-1}$ and the primed sum over $k$  means 
that for $m+s=0$ there is only one
term in the sum,
namely $k=0$, and for $m+s \neq 0$ the sum is over all 
the non-negative integers.
This is due to the fact that for $m+s=0$ we have 
the left ground state and $L_{-1}$
annihilates this state. The sum over $j$ takes care of the descendents coming 
from applying $J_-$ and finally sum over
$t$ takes into account the descendents coming 
from applying $G^i_{-\frac{1}{2}}$.
The upper bound on this sum means that if $m+s$ is less than 2, 
then we can only
apply a maximum of $m+s$  $G^i_{-\frac{1}{2}}$'s 
to the chiral primary. $d^A(t)$
takes into account the multiplicities of these 
descendents, together with the $Z_2$
actions. Since in model $II$, the $Z_2$ commutes 
with (4,4) supersymmetry, $d^{II}(t)
=d(t)$. On the other hand for models $I$ and $III$, 
the $Z_2$'s anticommute with 
$G^i_{-\frac{1}{2}}$'s and therefore $d^I(t)=d^{III}(t)=d(t)$ 
for $t$ even and
$d^I(1)=d^{III}(1)= -d(1)=2$.
Note that since $\epsilon^I(r,s) = (-1)^s$ the sum over $r$ yields 
zero on the right
hand side as expected for model $I$. For models $II$ and $III$, 
since $\epsilon(r,s)$
contains $(-1)^r$, the summation over $r$ gives a factor of 4. 
Equation (\ref{egsgrsp1}) reduces to:
\begin{eqnarray}
Z^{II}(p,q,y)&=& 
2 pq \frac{(1+p)^2}{(1-q)(y-y^{-1})}\left[y^3 \frac{(1-q^{1\over 2}y^{-1})^2}
{1-pq^{\frac{1}{2}}y}-y^{-3} \frac{(1- q^{1\over 2}y)^2}
{1-pq^{\frac{1}{2}}y^{-1}}\right]\nonumber\\ ~~&~& +
2p + 2\frac{p^2 +2p}{1-q}(q^{\frac{1}{2}}(y+ y^{-1})-2q)
\label{speg2}
\end{eqnarray}
\begin{eqnarray}
Z^{III}(p,q,y)&=&
2 pq \frac{(1-p)^2}{(1-q)(y-y^{-1})}\left[y^3 \frac{(1+ q^{1\over 2}y^{-1})^2}
{1-pq^{\frac{1}{2}}y}-y^{-3} \frac{(1+ q^{1\over 2}y)^2}
{1-pq^{\frac{1}{2}}y^{-1}}\right]\nonumber\\~~ &~&+
2p + 2\frac{p^2-2p}{1-q}(q^{\frac{1}{2}}(y+ y^{-1})+2q)
\label{speg3}
\end{eqnarray}

Expanding the above expressions in power series in $p^n q^m$ it is clear that  
$c^{A+}_{\rm sgr}(n,m,\ell)=0$
for $m < n/4$ except for $(n,m)=(1,0)$ and in this case
$c^{A+}_{\rm sgr}(1,0,0)=2$ which gives rise to a double pole in
(\ref{egsgrmp}) at $p=1$ together with a factor $(1+p)^2$ 
in the numerator. In fact
for $m > 0$ the only value of $m$ for which 
$c^{A+}_{\rm sgr}(4m,m,\ell) \neq 0$ is for $m=1/2$.

This means that the elliptic genus is of the
form 
\be
Z^A_{\rm multiparticle}(p,q,y) = 
\sum_{m,\ell} \frac{1}{(1-p)^2}P_{m\ell}(p)q^my^{\ell}
\ee
where $P_{m\ell}$ is a polynomial in $p$ of degree $4m+2$ (after
including the $(1+p)^2$ factor in the numerator of (\ref{egsgrmp})). We can
express 
\be
\frac{P_{m\ell}(p)}{(1-p)^2} = \frac{a_{m\ell}}{(1-
p)^2}+\frac{b_{m\ell}}{1-p} + P'_{m\ell}(p)
\ee
where $P'_{m\ell}$ is a polynomial of degree $4m$. This means that
if we restrict ourselves to $m < n/4$, then only the  
coefficients $a_{m\ell}$ and $b_{m\ell}$ contribute to the elliptic
genus. 

The same applies also to the CFT side. We start from the Ramond elliptic
genus for $\tilde{y}=1$ which is given by
\be
Z^A_{\rm cft}(p,q,y) = \prod_{n,m,\ell} \left[\frac{1+p^n q^m y^{\ell}}
{1-p^n q^m y^{\ell}}\right]^{c_{\rm cft}^{A+}(mn,\ell)}
\label{cfteg}
\ee
where we have used the fact that for $\tilde{y}=1$ the partition
function for the identity sector vanishes. 
$c_{\rm cft}^{A+}(m,\ell)$ are given by
$\frac{1}{2}{\rm tr}g^A q^m y^{\ell}$
where the supertrace is taken over the (4,4) CFT with target space $T^4$
in zero momentum and winding sectors. $g^{II}$ is $I_4$ and
$g^{III}$ is $(-1)^{F_L}I_4$. It follows that 
\begin{eqnarray}
\sum_{m,\ell}c_{\rm cft}^{II+}(m,\ell)q^m y^{\ell} &=& 8 \left[\frac{\theta_2(q,z)}
{\theta_2(q,0)}\right]^2 \nonumber\\
\sum_{m,\ell}c_{\rm cft}^{III+}(m,\ell)q^m y^{\ell} &=&- 8 \left[\frac{\theta_1(q,z)}
{\theta_2(q,0)}\right]^2 
\end{eqnarray}
where $y=e^{2\pi i z}$.
Using the fact that in both these models, the $U(1)$, which couples to $y$,
comes with a current algebra
which commutes with the $Z_2$ action and contributes to the total 
stress energy tensor via a Sugawara
term, we conclude that
\be
c_{\rm cft}^{A+}(m,\ell) \equiv c_{\rm cft}^{A+}(4m-\ell^2)
\ee
is only a function of $(4m-\ell^2)$. Furthermore 
$c_{\rm cft}^{A+}(4m-\ell^2)$ vanish when the 
argument is less than -1 and satisfy, for model $II$
\begin{eqnarray}
\sum_{\ell}c_{\rm cft}^{II+}(4m-\ell^2) &=& 8\delta_{m0},~~~~~~
\sum_{\ell} \ell c_{\rm cft}^{II+}(4m-\ell^2) =0,
\nonumber\\c_{\rm cft}^{II+}(0)
&=&2c_{\rm cft}^{II+}(-1)=4 ,
\label{cftcrel2}
\end{eqnarray}
and for model $III$
\begin{eqnarray}
\sum_{\ell}c_{\rm cft}^{III+}(4m-\ell^2) &=& 0,~~~~~~~~
\sum_{\ell} \ell c_{\rm cft}^{III+}(4m-\ell^2) =0,
\nonumber\\ c_{\rm cft}^{III+}(0)
&=&-2c_{\rm cft}^{III+}(-1)=-4
\label{cftcrel3}
\end{eqnarray}
 
In the NS sector the elliptic genus is obtained by the spectral flow
and is given by the same expression as (\ref{egsgrmp}), with the exponent
replaced by $c_{\rm cft}^{A+}(4mn-n^2-\ell^2)$. For both models, for $m > 0$
the latter are nonvanishing only for $n \leq 4m$ and for $m=0$, $n$ must be 1 
and $\ell=0$. For $m=\ell=0$ and $n=1$
using the above equation we find that in both the models the elliptic genus 
have double poles at $p=1$ and there is a factor 
of $(1+p)^2$ in the numerator. 
We can now use the same arguments as in the supergravity case to
conclude that for dimensions less than $N/4$ only contribution comes 
from the coefficients of
the double pole and the single pole at $p=1$.

Thus the matching of CFT states and supergravity states for
dimensions less than $N/4$ implies that the coefficients of the
double pole and the single pole at $p=1$ should match on the two
sides. This in turn implies
\bea
\sum_nc_{\rm cft}^{A+}(4mn-n^2-\ell^2)&=&\sum_n c_{\rm sgr}^{A+}(m,n,\ell)
\nonumber\\ \sum_n nc_{\rm cft}^{A+}(4mn-n^2-\ell^2)&=&\sum_n nc_{\rm 
sgr}^{A+}
(m,n,\ell) 
\eea

We first compute these quantities in the supergravity side. The two
quantities that we are interested in are just $Z^A$ and the first
derivative of $Z^A$ with respect to $p$ evaluated at $p=1$. From
(\ref{speg2}) (\ref{speg3}) after some algebra one can show that for model $II$
\begin{eqnarray}
Z^{II}(1,q,y) &=& -14 -2\frac{q^{\frac{1}{2}}(y+y^{-1})+2q}{1-q}
+ \frac{8}{1- q^{\frac{1}{2}}y} + \frac{8}{1-q^{\frac{1}{2}}y^{-1}}
\nonumber\\
\frac{\partial Z^{II}(p,q,y)}{\partial p}|_{p=1} &=& 2 + 
\frac{8 q^{\frac{1}{2}}y}{(1- q^{\frac{1}{2}}y)^2} + 
\frac{8q^{\frac{1}{2}}y^{-1}}{1-q^{\frac{1}{2}}y^{-1}}
\label{sgrres2}
\end{eqnarray}
and for model $III$ 
\begin{eqnarray}
Z^{III}(1,q,y) &=& 2 -2\frac{q^{\frac{1}{2}}(y+y^{-1})+2q}{1-q}
\nonumber\\
\frac{\partial Z^{III}(p,q,y)}{\partial p}|_{p=1} &=& 2
\label{sgrres3}
\end{eqnarray}

In order to evaluate these quantities on the CFT side we define $t$ and $u$
via $m=t/2$ and $\ell = t-2u$ with $t$ and $u$ non-negative integers
so that $c_{\rm cft}^{A+}(4mn-n^2-\ell^2)=c_{\rm cft}^{A+}(4u(t-u)-
(n-t)^2)$. We can now sum over $n$ using equations
(\ref{cftcrel2}),(\ref{cftcrel3}), taking
care of the ranges of $n$. For $|\ell| \geq 2$, the sum over $n$ from 1
to infinity includes all the terms in (\ref{cftcrel2}) and
(\ref{cftcrel3}), 
while for $|\ell|=1$, $n=0$ term is missing and for $\ell=0$, $n=0$ term as
well as $n=-1$ (for $m=0$) are missing. The result for model $II$ is\\\\\
\begin{tabular}{lll}
$~~~~~$ & $\sum_n c_{\rm cft}^{II+}(4mn-n^2-\ell^2)$ & $\sum_n n
c_{\rm cft}^{II+}(4mn-n^2-\ell^2)$ \\  
$|\ell|\geq 2~~~~~$& $8 \delta_{m,\frac{|\ell|}{2}}$&$ 8|\ell|
\delta_{m,\frac{|\ell|}{2}}$ \\
$|\ell|=1~~~~~$&$ 8 \delta_{m,\frac{1}{2}}-2$&$ 8\delta_{m,\frac{1}{2}}$\\
$\ell=0~~~~~$&$ 6\delta_{m,0} -4$&$2\delta_{m,0}$ \\
\end{tabular}\\\\\
and for model $III$\\\\\
\begin{tabular}{lll}
$~~~~~$ & $\sum_n c_{\rm cft}^{III+}(4mn-n^2-\ell^2)$ & $\sum_n n
c_{\rm cft}^{III+}(4mn-n^2-\ell^2)$ \\  
$|\ell|\geq 2~~~~~$& $0$&$ 0$ \\
$|\ell|=1~~~~~$&$ -2$&$ 0$\\
$\ell=0~~~~~$&$ 4-2\delta_{m,0}$&$2\delta_{m,0}$ \\
\end{tabular}\\\\\

Expanding the supergravity results (\ref{sgrres2}), (\ref{sgrres3}), 
one finds complete
agreement between supergravity and CFT for model $II$. For model $III$
however, we find that all the numbers agree except for $\ell=0$ and
$m > 0$ for which supergravity gives $-4$ while CFT gives +4. To see this
descrepancy, let us look at this case more closely. For $\ell=0$
\be
\sum_{n=1}^{\infty}c_{\rm cft}^{III+}(4mn-n^2)= \sum_{n \in {\cal
Z}} c_{\rm cft}^{III+}(4m^2-(n-2m)^2)-
c_{\rm cft}^{III+}(0)-\delta_{m,0}c_{\rm cft}^{III+}(-1)
\ee
For $m=0$ the $c_{\rm cft}^{III+}(0)=-4$ corresponds 
to the multiplicities of the 
ground states and this certainly agrees with supergravity since this
information already entered in the Poincare' polynomial. For $m>0$ CFT
still gives $c_{\rm cft}^{III+}(0)=-4$ since they depend only on the
combination $(4m-\ell^2)$, while the supergravity result indicates that 
it should be +4, which happens to be the value of 
$c_{\rm cft}^{II+}(0)$. A CFT with $U(1)$ current 
algebra which couples to $y$
certainly cannot have this property.  We have already observed 
this mismatch of $c_{\rm
cft}^{III+}(0)$ for $m=0$ and $m>0$, when
discussing the U-duality predictions for the 3-charge system. U-
duality says that the elliptic genus for model $III$ should be the
same as model $II$ with $p$ and $q$ exchanged. Let us see if the
supergravity result agrees with the model $II$ after exchanging $p$
and $q$. From equation (\ref{cfteg}), which describes the symmetric
product part of the CFT, 
we see that this is
\be
\prod_{n=1}^{\infty}\prod_{m=0}^{\infty}\prod_{\ell}
\left[\frac{1+p^n q^m y^{\ell}}{1-p^n q^m
y^{\ell}}\right]^{c^+(mn,\ell)}
\label{udualeg}
\ee
where $c^+({m,\ell})$ are given by
\begin{eqnarray}
c^+(0,\ell)&=&c_{\rm cft}^{III+}(-\ell^2), ~~~~~~~~~~{\rm for}\,\, m=0
\nonumber\\
c^+(m,\ell)&=&c_{\rm cft}^{II+}(4m-\ell^2), ~~~~~~~~~~{\rm for}\,\, m>0
\end{eqnarray}
In (\ref{udualeg}) we have also included the contribution of the
center of mass CFT $(R^4 \times T^4/I_4)$ for 
model $II$, which, after
exchanging $p$ and $q$ is only a function of $p$. This in fact gives
the $m=0$ terms in (\ref{udualeg}). By exchanging $p$ and $q$ in the
symmetric product part we would also have obtained terms in the
product with $n=0$, but these terms just reproduce the 
elliptic genus of the center of mass CFT for model $III$, 
which the supergravity side cannot see.
Therefore, we have dropped these terms.
$c^+(m,\ell)$ defined above has exactly the property that it is $-4$
for $m=\ell=0$ and +4 for $m=\ell^2/4 >0$. Indeed, by repeating
the previous analysis for the elliptic genus (\ref{udualeg}), we find
complete agreement with the supergravity result for model $III$.
We will make some comments on this remarkable fact in the
conclusions.

\section{Type I' supergravity on $AdS_3\times S^3$}

In this section we determine the supergravity spectrum  
for type IIB/$\Omega I_4$, further compactified on $AdS_3\times S^3$.
We do not have a CFT description of the D1/D5 system in 
this theory. 
In this paper we will not
attempt to find this directly, like in the models
discussed in the previous section, but rather 
use supergravity to extract informations about the D1/D5
system in this background.  
As already mentioned in the introduction, one expects 
this D1/D5 system 
to be related to the better understood D1/D5 
system in type IIB on $K3$.
To be more precise, the map between orbifold group generators and
charges in the two backgrounds is given in the following table:

$$
\begin{CD}
{\bf C}@>S>>{\bf D} @>T_{15}>>{\bf E}@> S >>{\bf F}\\
D_1      &&  F_1       &&   p_1     &&  p_1     \\
D_{12345}&& NS_{12345} && NS_{12345}&&D_{12345} \\
p_1      &&  p_1       &&   F_1     &&  D_1     \\
I_4      &&  I_4       &&(-)^{F_L} I_4&&\Omega I_4\\
\end{CD}
$$\\
{\it Table 3: IIB on $K3$ versus type IIB/$\Omega I_4$ }\\

For a single D5-brane, $Q_5=1$, since the D1 charge
is associated to the power of $p$ while the KK charge
is associated to the power of $q$ in the expansion
of the elliptic genus, table 3 implies that
a counting formula for excitations in the D1/D5 system in 
type IIB/$\Omega I_4$
can be obtained from the type IIB on $K3$ result after 
$p\leftrightarrow q$ exchange.
We are then led  to the following prediction for the
D1-D5-KK multiplicities in IIB/$\Omega I_4$:
\be
\tilde {Z}(p,q,y,\tilde{y})=y_-^2 \tilde{y}_-^2  
\tilde{Z}_{cm}(q,y,\tilde{y})
\, \tilde{Z}_F(p,y,\tilde{y}) 
\tilde{Z}_{sp}(p,q,y,\tilde{y})
\label{ztilde}
\ee
where
\bea
\tilde{Z}_{cm}(q,y,\tilde{y})&=&{1\over \hat{\vartheta}_1(y \tilde{y}|q)
\hat{\vartheta}_1(y \tilde{y}^{-1}|q)\eta^{18}(q)}\nonumber\\
\tilde{Z}_{sp}(p,q,y,\tilde{y})&=& \prod_{n,m=1}^{\infty}
(1-p^n q^m y^l \tilde{y}^{\tilde{l}})^{-c_{K3}(nm,l,\tilde{l})}
\eea
both come from the symmetric product piece in 
the dual $R^4\times (K3)^N/S_N$ and
\be
\tilde{Z}_{F}(p,y,\tilde{y})=\frac{\hat{\vartheta}_1^2(\tilde{y}|p)}
{\hat{\vartheta}_1(y \tilde{y}|p)
\hat{\vartheta}_1(y \tilde{y}^{-1}|p)}
\label{zftilde}
\ee
together with the zero mode factor $y^2_- \tilde{y}^2_-$ is the 
contribution of the center of mass $R^4$. 
Notice that unlike in the more familiar cases of type IIB on $T^4$
or $K3$, $Z_F(p,y,\tilde{y})$ is not the naive extension 
to $p=0$ of $\tilde{Z}_{sp}(p,q,y,\tilde{y})$ and therefore
(\ref{ztilde}) does not admit a description in terms of a 
CFT partition function. We can however compare (\ref{ztilde})
with the prediction from supergravity on $AdS_3\times S^3$,
in following the logic of the previous section for the freely
acting orbifold model $III$.

The study of chiral primary states in the 
AdS reduction of the present supergravity theory 
follows closely our analysis
for model $III$ in the previous section, with some  
important differences. First of all, the  $(4,0)$ supermultiplets with
negative eigenvalue under $\Omega I_4$ are now projected out   
already at the level of one-particle states. This should be contrasted
with the case where the $\Omega I_4$ orbifold group action is
accompained with a shift and supermultiplets with both
eigenvalues enter in the computation although counted with
different signs. Indeed, in the absence of a shift, the boundary
theory is no longer a $Z_2$ orbifold of the parent one in type IIB 
on $T^4$, and looks (even locally) drastically different from it.
The second important difference is the presence in this case
of open string sectors living on
D5-branes. They lead to 16 additional vector multiplets of
the ${\cal N}=(1,1)$ six-dimensional supersymmetry. The (4,0)
multiplet content together with degree for the vector multiplets is
given in (\ref{vI}).

To keep our expressions as compact as possible and our discussion
close to our previous analysis, we will insist in organizing
states in terms of ``$(4,4)$ supermultiplets'', with $(4,0)$
supermultiplets inside them taken with alternate $\Omega I_4$
eigenvalues. The restriction to states with even $\Omega I_4$
eigenvalues will be performed only at the end by keeping
only descendants with even(odd) numbers of $G_{-1/2}^i$
acting on a given even(odd) chiral primary ground state.
In fact in (\ref{vI}) the vector multiplets are already expressed
in terms of odd (4,4) multiplets. Using eq.(\ref{vI}) for 
$n_V-4$ vector
multiplets (in the present case $n_V=20$)
living on D5-branes together
with the results for untwisted states from 
model $III$, we are then left 
with the one-particle supergravity Hilbert space
\be
{\cal H}^A_{\rm single~particle}=\oplus'_{m \geq 0} \,
\tilde{h}_{r,s}\,
({\bf m+r},{\bf
m+s})^{\epsilon_{III}(r,s)}_{m+1}  
\label{shortI'}
\ee
with
\be
(-1)^{r+s}\tilde{h}_{r,s}=d(r)\left[d(s)-\frac{n_V-4}{2} \delta_{s,1}\right]  
\ee
and $\epsilon_{III}=(-)^s$. As before, we read off the spectrum 
of (chiral, chiral) primary states from the supergravity 
Poincare' polynomial, which after keeping even chiral primaries 
in (\ref{shortI'}) reduces to:
\be
P_{\infty}= \frac{(1-y)^2}{(1-y^2)(1-\tilde{y}^2)(1-y\tilde{y})} 
\prod_{m=1}^{\infty}\frac{(1-y^{m+1}\tilde{y}^m)^2 (1-y^m\tilde{y}^{m+1})^2}
{(1-y^{m+1}\tilde{y}^{m+1})^2 (1-y^{m+2}\tilde{y}^m)(1-y^m\tilde{y}^{m+2})}
\ee 
where we have added the usual missing multiplets $2(1,0)+(2,0)+(0,2)$.
One can easily see that this agrees with the residue of the first 
order pole at $p=1$ of (\ref{zftilde}), after a spectral flow to the
NS sector.

The spectrum of descendants of the form (chiral, anything) is 
now determined from
\bea
\tilde{Z}(p,q,y) =\sum_{m,r,s}
\sum_{t=0}^{min(2,m+r)} {\sum_{k=0}^{\infty}}' \tilde{h}^{r,s}(-)^s
\left({1+(-)^{t+s}\over 2}\right)
p^{m+1}q^{\frac{m+r+t}{2}+k} \sum_{j=0}^{m+r-t}y^{m+r-t-2j}    
\nonumber
\eea
where the projector $\left(\frac{1+(-)^{t+s}}{2}\right)$ has been inserted
in the trace in order to ensure that only states with 
$\Omega I_4$ eigenvalue +1
enter in the sum.
After a straightforward algebra, one finds 
\bea
\tilde{Z}(p,q,y) &=&
2 pq \frac{(1-p)^2}{(1-q)(y-y^{-1})}
\left[\frac{y^3+y q+\frac{n_V}{2} y^2 q^{1\over 2}}
{1-pq^{\frac{1}{2}}y}-\frac{y^{-3}+q y^{-1}+\frac{n_V}{2}q^{1\over 2}y^{-2}}
{1-pq^{\frac{1}{2}}y^{-1}}\right]\nonumber\\~~ 
&~&+
2p + 2\frac{p^2-2p}{1-q}(q^{\frac{1}{2}}(y+ y^{-1})+\frac{n_V}{2}q)
\eea
In particular
\begin{eqnarray}
\tilde{Z}(1,q,y) &=& 2 -\frac{2}{1-q}\left[q^{\frac{1}{2}}(y+y^{-1})+
\frac{n_V}{2}q\right]
\nonumber\\
\frac{\partial \tilde{Z}(p,q,y)}{\partial p}|_{p=1} &=& 2
\label{sgrresI'}
\end{eqnarray}
leading to\\

\begin{tabular}{lll}
$~~~~~$ & $\sum_n c_{\rm sugra}^{I'}(m,n,l)$ & $\sum_n n
c_{\rm sugra}^{I'}(m,n,l)$ \\  
$|\ell|\geq 2~~~~~$& $0$&$ 0$ \\
$|\ell|=1~~~~~$&$-2$&$0$\\
$\ell=0~~~~~$&$ (n_V+2)\delta_{m,0} -n_V$&$2\delta_{m,0}$ \\
\end{tabular}\\\\\

One can easily see that these results for $n_V=20$ are in complete
agreement with $\sum_n c_{\rm CFT}(m,n,l)$ and 
$\sum_n n c_{\rm CFT}(m,n,l)$, with $c_{CFT}(n,m,l)$ the 
coefficients coming from the $K3$ symmetric product expression 
after $p\leftrightarrow q$ exchange i.e. 
$c_{CFT}(m,l)=c_{K3}(m.l)$ for $m\neq 0$;
$c_{CFT}(0,0)=-2c_{CFT}(0,\pm 1)=-4$.

Therefore supergravity agrees with U-duality even in this case.

\section{Conclusions}

In this paper we studied the 6-dimensional (2,0) and (1,1)
supergravity
theories on $AdS_3 \times S^3$ arising from orbifolding/orientifolding 
of IIB theory
on $T^4$. These theories arise in certain type
II or  type I compactifications. While (2,0) theories give rise to
(4,4) superconformal symmetry $SU(1,1|2)_R\times SU(1,1|2)_L$, the (1,1)
theories 
only have (4,0)
symmetry $SU(1,1|2)_R\times SU(1,1)_L\times SU(2)_L$.  
In the case of freely acting orbifolds the Poincare' polynomials 
encoding the multiplicities of (chiral,chiral) primaries in the
AdS supergravity theories were computed and shown to agree
with the predictions coming from 
D1/D5 CFTs proposed in \cite{ghmn}. 
We also computed
the supergravity Poincare' polynomial for type I', which agrees with the 
pure D1/D5 bound state multiplicities as predicted by U-duality.
We extended the supergravity analysis to the elliptic genus,
which should encode the information about the multiplicities of
D1/D5/KK
bound states. In the
(4,4)
case this agrees with the elliptic genus for the (4,4) symmetric
product
CFT in the expected range of validity, namely $P < Q_1/4$, where 
$P$ is the KK momentum. 
On the other hand, in the (4,0) case
the supergravity elliptic genus does not agree with the one for the
symmetric product (4,0)  CFT proposed in \cite{ghmn}. 
Instead it agrees, in the
range of validity, with the 
multiplicities obtained from the elliptic genus of the U-dual (4,4)
CFT via $P \leftrightarrow Q_1$ exchange. Therefore, quite remarkably,
the (4,4) CFT
is probed by gravity duals in both $P < Q_1/4$ as well as
$Q_1 < P/4$ ranges; in the former range the gravity dual being the
usual (4,4) one, while in the latter it is the (4,0) $AdS_3$ supergravity.
The results therefore provide a very strong evidence for the
correctness of the proposed (4,4) symmetric product CFTs.

Although we have focussed here on the models considered in \cite{ghmn}
and the standard type I' model, one can easily generalize the methods 
given here to other models. To give an example, consider 6-dimensional
(2,1) supergravity obtained by comactifying IIB on $T^4/Z_2$, where the
$Z_2$ is acting asymmetrically only on the left movers together with a
shift in a transverse direction. We can do the KK analysis on
$AdS_3\times
S^3$ exactly as in section 3. The resulting spectra organize
themselves in (4,4) multiplets as in eq.(\ref{shortrs}), with the $Z_2$
action  being given by
$(-1)^r$. We can go on to compute the elliptic genus, and in the
regime of validity, the two relevant moments of expansion coefficients
are
given by eq.(\ref{sgrres3}), with the only difference being that the 
sign of the $2q$ term in the numerator
of the right hand side of the first equation is reversed. The
(4,4) symmetric product CFT which agrees with this supergravity is the
one corresponding to $(T^4)^N/S_N$ modded by the diagonal $Z_2$ 
action which acts on each copy of $T^4$ asymmetrically only on the 
right-moving sector.

However several important questions remain to be answered. For
example, what is the
correct (4,0) boundary CFT which is dual to the (4,0) $AdS_3$
supergravity? The analysis reported here shows that this CFT
can not be of a symmetric product type. In \cite{ghmn} it has 
been shown that there is no (4,0) CFT which completely reproduces the
elliptic genus of the U-dual (4,4) symmetric product CFT after $P
\leftrightarrow Q_1$ exchange. What we are asking here, however, is 
a weaker condition, namely, we are looking for a (4,0) boundary CFT
which reproduces the elliptic genus of the (4,0) $AdS_3$ supergravity
in the region of validity. That such a CFT should exist follows from
the AdS/CFT correspondence.

In \cite{ghmn}, using the same adiabatic type arguments employed to
derive (4,4) CFT for freely acting $Z_2$ orbifold,  
we had obtained certain (4,0) symmetric product CFTs
for freely acting orientifold models. These (4,0) CFTs gave the
correct ground state multiplicities, however they were not in
agreement with U-duality for excited states, if one assumes that the
symmetric product CFT is defined for vanishing RR fields. 
However, we argued there that the symmetric product CFT
is sitting at 
$\chi=1/2$  component of the moduli space, where $\chi$ is the RR
0-form. In model $II$ or in IIB on $K3$ (i.e. (4,4) models), $\chi$
is a modulus and therefore one can go continuously from $\chi=1/2$ to
$\chi=0$ point. This is not the case for the (4,0) models appearing in
the model $III$ and type I'. In these cases $\chi$ is projected out
and,
as a result, $\chi=0$ and $\chi=1/2$ define two disconnected components
of the moduli space.  In \cite{ghmn}, we presented an example to show
that the physics of these two different components, such as the number of
BPS states, can in general be
very different. Now U-duality maps the $\chi=1/2$ point of the
(4,4) theory to the $\chi=0$ component of the (4,0) theory, and not to
the $\chi=1/2$ component, where the symmetric product CFT exists. This
could possibly explain the fact that the two symmetric product CFT~s
for (4,4) and (4,0) models did not match under U-duality map. The
supergravity analysis, we have presented here, is presumably valid at 
$\chi=0$. For the (4,4) case, as pointed out above, the physics at
$\chi=0$ should be the same as at $\chi=1/2$, and therefore, the
supergravity analysis agrees with the (4,4) symmetric product CFT. On
the other hand, for the (4,0) case, there is no reason to expect that
the elliptic genus for these two different components should be the
same.  In
fact, from what has been said above, it is clear that the (4,0)
supergravity 
elliptic genus at $\chi=0$ should be related by
U-duality to that of the (4,4) symmetric product CFT, as we indeed
found in this paper. If this is
the explanation for the apparent discrepancies, we must still
look for the supergravity duals of the (4,0) symmetric product CFT~s
of
\cite{ghmn}, whose derivation was at the same level of
rigour as the derivation of (4,4) models. This would, in particular, 
involve an understanding of how
$\chi=1/2$ background could modify the supergravity (or more precisely
superstring on $AdS_3$) analysis.

\vskip 0.5in
{\bf Acknowledgements}

We acknowledge discussions with J. de Boer.
This project is supported in part by EEC under TMR contracts
ERBFMRX-CT96-0090, HPRN-CT-2000-00148 ,
HPRN-CT-2000-00122 and the INTAS project
991590. The work of J.F.M. was partially supported by the INFN
section of University of Rome ``Tor Vergata''.

\rnc{\Large}{\normalsize}


\begin{thebibliography}{00}
\addcontentsline{toc}{section}{References}
\frenchspacing
\small
\addtolength{\itemsep}{-4pt}


\bibitem{maldacena} J.M. Maldacena,
{\it The Large N Limit of Superconformal Field Theories and Supergravity},
Adv.Theor.Math.Phys. 2 (1998) 231; Int. J. Theor. Phys. {\bf 38} (1999) 1113,
 hep-th/9711200.

\bibitem{V} C. Vafa, {\it Instantons on D-branes}, Nucl. Phys. {\bf B463}
(1996) 435, hep-th/9512078.

\bibitem{SV} A. Strominger and  C. Vafa,
{\it Microscopic Origin of the Bekenstein-Hawking Entropy},
Phys. Lett. {\bf B379} (1996) 99, hep-th/9601029.

\bibitem{W1} E. Witten, {\it On the Conformal Field Theory 
of the Higgs Branch}  JHEP {\bf 07} (1997) 003, hep-th/9707093.

\bibitem{GKS} A. Giveon, D. Kutasov and N. Seiberg, {\it Comments on
String Theory on $AdS_3$}, Adv. Theor. Math. Phys. {\bf 2} (1998) 733,
hep-th/9806194.  

\bibitem{D1} R.Dijkgraaf, {\it
Instanton Strings and Hyperkhaler
Geometry}, Nucl. Phys. {\bf B543} (1999) 545,
hep-th/9810210.


\bibitem{MS} J.M. Maldacena and A. Strominger,
{\it AdS3 Black Holes and a Stringy Exclusion Principle},
JHEP {\bf 9812} (1998) 005, hep-th/9804085.


\bibitem{DB1} J. de Boer, {\it Six-Dimensional Supergravity on
$S^3\times AdS_3$ and 2D Conformal Field Theory}, 
Nucl. Phys. {\bf B548} (1999) 139, hep-th/9806104. 

\bibitem{DB2} J. de Boer, {\it Large N Elliptic Genus and AdS/CFT
Correspondence}, JHEP {\bf 9905} (1999) 017, hep-th/9812240.

\bibitem{mms} J.M. Maldacena, G. Moore and  A. Strominger,
{\it Counting BPS Black Holes in Toroidal Type II String Theory},
hep-th/9903163.


\bibitem{ghmn} E. Gava, A.B.  Hammou, J.F. Morales, K.S. Narain,
{\it The D1/D5 System in ${\cal N}=4$ String Theories}, hep-th/0012118.

\bibitem{dmvv} R. Dijkgraaf, G. Moore, E. Verlinde, and H. Verlinde, 
{\it Elliptic Genera of Symmetric Products and Second Quantized Strings }, 
Comm. Math. Phys. {\bf 185} (1997) 197.
  
\bibitem{SS} A. Salam and J. Strathdee, {\it On Kaluza-Klein Theory},
Ann. Phys. {\bf 114} (1982) 316.

\bibitem{dkss} S. Deger, A. Kaya, E. Sezgin and P. Sundell, {\it
Spectrum of D=6, N=4b Supergravity on $AdS_3 \times S^3$},
Nucl. Phys. {\bf B536} (1998) 110, hep-th/9804166.

\bibitem{KS} D. Kutasov and N. Seiberg, {\it More Comments on String
Theory on $AdS_3$}, JHEP {\bf 9904} (1999) 008, hep-th/9903219.



\end{thebibliography}
\end{document}